# Comparison of Three Job Mapping Algorithms for Supercomputer Resource Managers

A. V. Baranov, E. A. Kiselev, B. M. Shabanov, A. A. Sorokin, and P. N. Telegin

**Abstract**—Performance of supercomputer depends on the quality of resource manager, one of its functions is assignment of jobs to the nodes of clusters or MPP computers. Parts of parallel programs interact with each other with different intensity, and mapping of program to supercomputer nodes influence efficiency of the run. At each program run graph representing application program is to be mapped onto graph of nodes representing a subset of computer system. The both graphs are not known beforehand, hence the mapping must be done in reasonable time while scheduling resources. Three mapping algorithms were explored: parallel versions of simulated annealing, genetic and composite algorithms. A set of experimental runs with different algorithms parameters was performed, comparison of mapping quality and runtime was made, and suggestions on applicability of algorithms for resource managers were provided.

## 1. INTRODUCTION

A supercomputer today consists of computing nodes connected by a high-speed network. The network topology can vary and includes communication channels with different bandwidth and latency. This way a supercomputer can be represented as a graph $G_s = (B, E_s)$. $B$ is a set of vertices representing computational nodes. $E_s$ is a set of edges representing communication channels between the nodes. The throughput of the communication line is characterized by the weights of the edges of the graph $m_{ij}, (i,j = 1...N), (i,j) \in E_s$.

The main purpose of a supercomputer is the execution of parallel application programs. A parallel program is a set of processes that exchange information with each other with different intensity. A parallel program is represented by a graph $G_p = (A_p, E_p)$, where $A_p$ is a set of vertices corresponding to processes, $E_p$ is a set of edges, it represents information links between processes. Number of processes is $N = |A_p|$. The weights $c_{ij}, (i,j = 1...N)$ are assigned to the edges of the graph, The weights are reflecting the intensity of information exchange between the processes $i$ and $j$. Let's name this graph a program graph or information graph.

Mapping of information graph of a parallel program $G_p$ onto supercomputer structure graph $G_s$, is denoted as $\phi : A_p \to B$. The mapping is represented by matrix $X = \{X_{ij} : i \in A_p, j \in B\}$, where $X_{ij} = 1$, if $\phi(i) = j$, and $X_{ij} = 0$, otherwise. Assume that all computational nodes are of equal performance. In this case the criterion of the mapping optimality is the functional $F(X)$ [1]:

$$F(X) = \sum_{i=1}^{N}\sum_{j=1}^{N}\sum_{p=1}^{N}\sum_{k=1}^{N} m_{ij} c_{kp} X_{ki} X_{pj} \longrightarrow \min \qquad (1)$$

The problem has to be is solved by supercomputer resource managers. The manager receives a stream of user jobs, submitting them in a queue. Each job includes a parallel program that requires a number of supercomputer nodes for execution. When a job is launched, a subset of free nodes is allocated, i.e. it is not known in advance which specific nodes will be allocated for the job. In accordance with (1), for each job, we have to find the optimal mapping of the program graph onto a graph of a subset of selected nodes, which is not known beforehand. Time required by mapping mush be reasonable, i.e. it should not exceed this time of system timeouts (up to 5–15 minutes). Note that it is much less than the average job execution time of several hours.

In [2], a two-stage PGA (Partition-Genetics-Annealing) method for mapping a program graph onto a supercomputer structure graph is proposed for the resource manager of the supercomputer MVS-1000 [3, 4]. At the first stage, when the job is launched, the supercomputer nodes are selected from the set of free nodes. The selection of nodes is done using a modified algorithm for finding the min-cut partitioning of a graph [5]. This allows to select the subset of the most tightly coupled nodes from the set of free ones. At the second stage, the program graph is directly mapped onto the graph of the selected nodes.

To find the mapping, selected nodes are used. A parallel annealing simulation algorithm with genetic operations PAG (Parallel-Annealing-Genetics) on these nodes runs before starting the job. The parallel algorithm is based on a combination of simulated annealing and a genetic algorithm, his made it possible to obtain high-precision mapping in a short time. The PAG algorithm was tested on the MVS-1000 supercomputer, and for a number of applications it was possible to obtain a performance increase of up to 45%. The MVS-1000 consisted of only 32 single-processor nodes.

In this paper, we continue the study of algorithms for mapping the information graph of a parallel program onto the graph of a parallel computer.

We considered the ways to improve the solution [2] and studied the features of the composite parallel algorithm (imitation of annealing and genetic algorithm) for graphs of higher order corresponding to highly parallel computer.

## 2. RELATED WORK

To find the optimal mapping, exact algorithms can be applied which always find the minimum value of the objective function. Exact algorithms include a brute-force search of all possible mappings and the branch-and-bound method. Since the mapping problem is an NP-complete one, the running time of exact algorithms for graphs of medium and large sizes can be unacceptably long. Reducing the time to acceptable values is possible by the use of approximate parallel algorithms. Among approximate algorithms, there are heuristic ones. Heuristic algorithms can be iterative and population based algorithms. The iterative ones include the taboo search algorithm, the greedy random adaptive search algorithm, and the simulated annealing algorithm. Population algorithms include genetic algorithms, the particle swarm and the ant colony optimization algorithms.

The article [6] compares 11 heuristics used for mapping. Genetic algorithm provides the most accurate solution. The simulated annealing algorithm finds the solution in the shortest time, but the mapping accuracy is lower. The reason for the poor accuracy is the following. At the initial stages of operation, the simulated annealing algorithm can make unsatisfactory decisions, which it will be difficult to correct at later stages.

In [7], the authors developed an algorithm based on the ant colony optimization algorithm. It provides a solution on average 16% better than other heuristics. The simulated annealing algorithm typically finds a worse solution, having the minimum average running time.

In [8], the authors use a parallel version of the simulated annealing algorithm to solve the mapping problem. For each node, an initial solution is generated with a timestamp labelled as zero. Each thread at each node generates a solution based on the temporary global optimum solution. At each node there is a master thread that updates the solution within the node. Then it attempts to update the overall solution. Then all nodes generate new solutions based on an updated one.

The article [9] notes another disadvantage of the simulated annealing algorithm. It is examination of huge number of potential solutions. The authors propose a greedy strategy for generating an initial mapping. Jobs are sorted according to their communication requirements. A certain percentage of jobs from the top of the sorted list is assigned to adjacent nodes; all other jobs are assigned randomly. At the generation of a new solution it is attempted to place vertices with a large number of communications next to each other.

The article [10] presents a two-stage sequential simulated annealing algorithm. The calculation of the objective function for the new solution is represented as the objective function of the solution from which the candidate one is derived, plus the difference that is added by making changes to the display. The authors introduce the concept of quality of the assignment, where quality is the ratio of the value of the objective function of an optimal mapping to the value of the objective function of the resulting acceptable mapping.

In [11], the authors developed another simulated annealing method that can "correctly" determine the initial and final temperatures and the number of iterations needed for each temperature level, in order to eliminate unnecessary iterations.

The paper [12] presents a comparative analysis of use the simulated annealing algorithm for the mapping problem. The authors present optimal move functions for generation of a new mapping, compare acceptor functions, describe methods for setting the initial and final temperatures, and a method for choosing the number of iterations for given initial parameters of the algorithm.

The authors in [13] note that for a mapping search, the best result is achieved by using a genetic algorithm, but the algorithm runtime time is longer than that of other algorithms.

In the article [14], the authors compare the results of mapping search using a genetic algorithm and linear programming methods. The genetic algorithm, unlike linear programming, can find an acceptable solution in an acceptable time.

The article [15] considers the performance of a genetic algorithm on the graph partition problem compared to other standard heuristics. It is stated that the standard operators of the genetic algorithm are not sufficient to get a performance increase. In order to improve performance, a local the Kernighan-Lin [16] search algorithm is used, which makes it possible to find the best nearest solutions. The modified version of the genetic algorithm performs better than the simulated annealing and multiple local search algorithms.

The article [17] describes a parallel genetic algorithm improved by the hill climbing algorithm, which enables limitation of the solution space. It is possible to get the best mapping only with a large population size.

The authors of [18] developed a parallel genetic algorithm that splits populations between processes. Periodically, processes exchange the best members of the population and remove the worst ones. To minimize the time of communications, only neighboring processes exchange the best solutions, this allows achieving superlinear acceleration of the algorithm.

In [19], the authors use a genetic algorithm to solve the mapping problem for multiprocessor system-on-chip. They propose an approach where routing is evaluated as a fitness function, in contrary to the classical analytical evaluation, in which the main optimization indicators are the maximum packet transition delay and the average total packet delay.

In the article [20], for solving the mapping problem with a genetic algorithm, two functions for evaluating the quality of the mapping are introduced: penalty for computation imbalance and penalty for communication. The crossover and mutation operations are used. The article [21] describes the delay model using as the sum of the execution delay and the communication delay. There are the following restrictions on the proposed model: data moves only along the shortest path, only one other node interacts with any node at any time. During the execution of the algorithm, the system delay is first minimized, then vertices are bound to specific nodes. Crossover and mutation operations are used.

In [22–27], the authors describe the most common used computational model "Master-Slave" for a parallel genetic algorithm. "Master" generates new members of the population. All "Slave" processes only quantify the objective function value for new members of the population. This model of parallelism fits well on shared memory systems, as there is no time spent on communication interaction. The time of transferring new members of the population between computational nodes can exceed the time of the objective function value calculation. This will result in the inefficiency of this model in distributed memory systems.

In [22, 24–28], the authors also describe a parallel genetic algorithm with a ring exchange scheme. At each iteration of the ring transfers, different populations exchange the best members, transferring the best features between populations. The number of transmitted members of the population should not be large, as this can lead to similarity of the populations at the initial iterations of the communication exchange.

## 3. EXPLORED PARALLEL MAPPING ALGORITHMS

In our study, we implemented and compared three heuristic parallel algorithms: simulated annealing, genetic algorithm, and a composite algorithm. We used UGR-Metaheuristics [29] library to implement the algorithms.

The simulated annealing algorithm is based on the imitation of the physical process that occurs at the annealing of metals. The essence of the algorithm is the following.

1. A starting solution is generated. It becomes the candidate one.

2. A new solution is generated by swapping two arbitrary elements of the $X$ matrix.

3. If, after the permutation, the increment of the value of the functional (1) $\triangle F(X) < 0$, then the new solution becomes the candidate one. If $\triangle F(X) > 0$, then the new solution becomes the candidate one with the probability determined by the acceptor function. Accepting depends on the current annealing temperature, which decreases in the course of the algorithm.

4. The temperature of the system is reduced according to the temperature decrease function.

5. The algorithm stops when one of the values is reached: the certain number of iterations, the final temperature of the system, the number of consecutive iterations without improving the value of the objective function. Otherwise, go to step 2.

In the algorithm, several processes (threads) search for a solution. The best found candidate solution is broadcasted to all processes (threads), and each of them makes the received solution the candidate one. Parallel simulated annealing allows investigate more mapping options from the solution space than the sequential algorithm in the same time. Due to this, the obtained candidate solutions cover a greater number of local minima of the objective function. Thus, the probability of finding a better solution increases.

In terms of the genetic algorithm, an individual is represented by an array $p$, $i^{th}$ element of which (gene) contains the number of the node to which the $i^{th}$ process of the parallel program will be assigned. Individuals make up a population, its size is equal to or greater than the number of vertices of the program graph. The crossover operation exchanges genes between two individuals of a population. The mutation operation with a given probability changes a given number of individuals. The selection operation selects individuals from the population with the lowest value of the functional (1).

The parallel genetic algorithm is based on the ring information exchange scheme. Each process performs the following steps.

1. Generation of an initial population based on a pseudo-random sequence and setting thepopulation as the candidate one.

2. Breeding new descendants by applying the crossover operation to selected individuals of the current population.

3. Applying the mutation operation to the descendants generated in the previous step.
4. Replacing the worst solutions in the population with new descendants.

5. Selection of the best member of the population.

6. Communication exchange of the best solutions between the neighbour processes.

7. Replacing the worst solution in the population with a new solution, if it is better.

8. If the specified number of iterations was not been reached, go to step 2.

9. Selection of an individual with the minimum value of the objective function (1) among allprocesses. The selected individual is accepted as a solution to the mapping problem.

Unlike the case of simulated annealing algorithm, the generated solutions in genetic algorithms are quite different from the initial one. Due to this, it is possible to cover more local minima of the objective function and get closer to the global minimum. The advantage of the simulated annealing algorithm is that it takes less time to find an acceptable solution, while the advantage of the genetic algorithm is obtaining a more accurate solution. It seems reasonable to combine the two algorithms in order to provide high mapping accuracy in a reasonable time. Let's consider our proposed composite parallel algorithm.

At the first stage of the composite algorithm, a parallel simulated annealing algorithm works. In contrast to the simulated annealing discussed above, our algorithm does not make exchanges between processes. Each process generates a given number of solutions, which become the initial population for the parallel genetic algorithm. The choice of operation types of the genetic algorithm is made in accordance with the given configuration.

The parallel composite mapping search algorithm consists of the following steps:

1. Parallel search for solutions by each process using simulated annealing.

2. Generation of a population of solutions for the parallel genetic algorithm from the solutionsform step 1.

3. Running a parallel genetic algorithm for a given number of iterations.

4. Choosing the best solution in each process.

5. Choosing the best global solution.

The absence of exchanges at the stage of the parallel annealing simulation algorithm makes each process to generate a unique population of solutions. Due to the migration of solutions, it is possible to transfer the best features between populations, which will be inherited by descendants.

## 4. EXPERIMENT METHOD

The mapping search was processed for the input stream of different jobs coming to the supercomputer queue from different users. The subset of supercomputer nodes and, hence, the graph of the computing system is known only at the moment of the job (parallel program) start.

The parallel simulated annealing algorithm, the genetic algorithm, and the composite algorithm were implemented as an application program. Experimental runs were performed on the Broadwell section of the MVS-10P OP supercomputer [30, 31] at the JSCC RAS. The section consists of 136 nodes with the following features:

- 2 Intel Xeon E5-2697Av4 processors;

- 32 physical, 64 virtual cores in the node;

- 128 GB of RAM;

- Intel Omni-Path interconnect.

A set of taiXeyy [1, 32] graphs of various orders was used as workload. X denotes the order of graph, and yy is a version of graphs with a given order. A graph of a computer system and an information graph of the program are specified for each number of vertices as the distance matrices. For each pair of graphs used in the experiments, the distance matrices and the minimum value of the objective function (1) are known. The taiXeyy set is used by researches to test mapping algorithms for quadratic assignment problems.

We used the following instances in experiments: tai27e01, tai45e01, tai75e01, tai125e01, tai175e01, tai343e01, tai729e01. The experiment consisted in finding a mapping for a given pair of graphs with fixed algorithm parameters and the number of processes. To average the results of 10 runs were performed with the same parameters for each experiment.

At the first stage of the experiments the selection of optimal algorithm parameters was done. At the second stage, an experimental comparison of the algorithms parameters with the found optimal parameters was performed.

## 5. EXPERIMENTAL RESULTS

The following parameters affect the convergence of the parallel annealed simulation algorithm:

- number of examined solutions at a fixed temperature;
- number of successfully found solutions at a given temperature;
- temperature decrease function;
- number of successive iterations of the algorithm;
- number of iterations of the parallel algorithm;
- the number of "solvers" in one process;
- parameters for initial temperature setting.

Values suggested by the author of the UGR-Metaheuristics [29] library were used for the parameters that influence the number of found solutions at single temperature and the setting of the initial temperature.

The experiment consisted in multiple mapping search for graphs of a given order with a different number of branches of parallel program and fixed algorithm parameters. Only the value of the selected parameter was changed.

For a fixed temperature, the simulated annealing algorithm examines several candidate solutions. The number of examined solutions is set by the value of the maxNeighbors parameter. The tai343e01 instance representing graphs consisting of 343 vertices was chosen as a test dataset. The results of selecting the value of the maxNeighbors parameter are presented in Figure 1.

The dependence of the mapping search time on the number of examined variants is given in Figure 2.

The best average value of the objective function is achieved with value of parameter maxNeighbours = 50. In this case, the algorithm has an acceptable average mapping search time.

Another important parameter is the temperature decrease function. We considered two temperature decrease functions implemented in the UGR-Metaheuristics library: the linear function $T_{next} = q \times T$ and the Cauchy function $T_{next} = \frac{T}{1+\beta \times T}$. Coefficient $\beta$ is calculated by the following formula

$$\beta = \frac{T_{final} - T_{init}}{\frac{M}{N} \times T_{final} \times T_{init}}$$

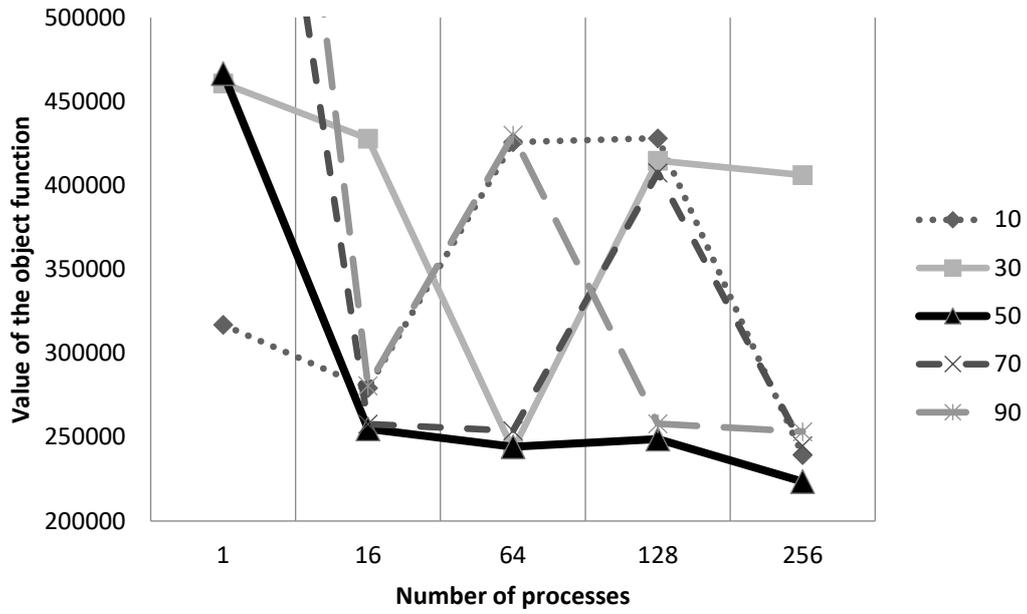

Figure 1. Values of the objective function for different values of maxNeighbours parameter

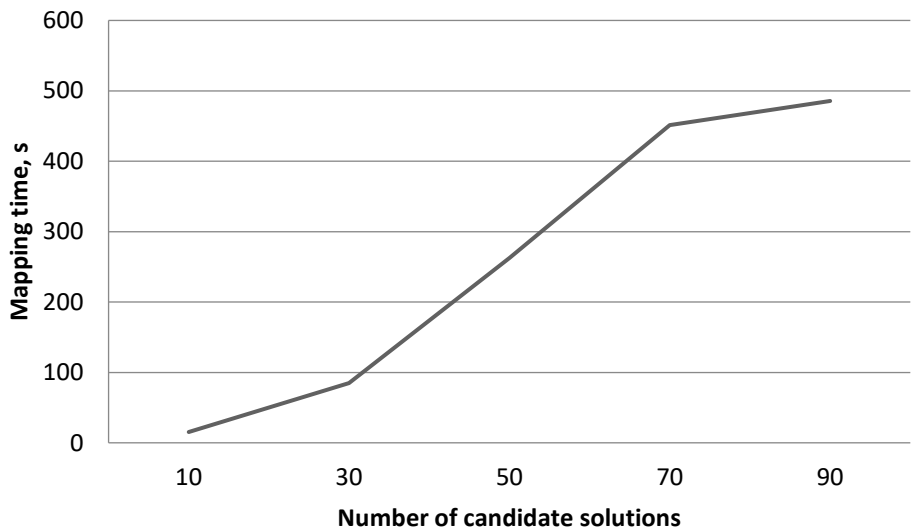

Figure 2. Mapping search time for different values of the maxNeighbors parameter

Results are shown in Figure 3.

The Cauchy function provides an average minimum execution time and an average minimum objective function value.

1600000

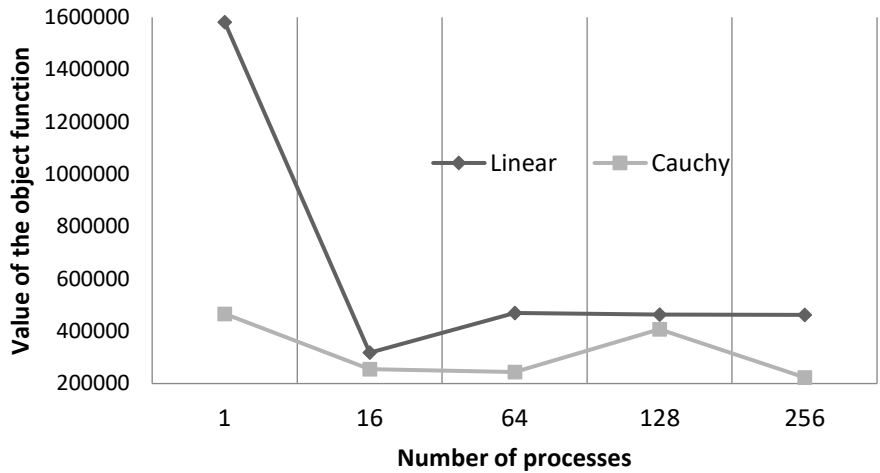

Figure 3. Dependence of objective function values on the temperature function

The parallel algorithm is in the simultaneous operation of several processes, each of which performs a certain number of successive iterations of the simulated annealing algorithm. After performing this number of iterations, the processes exchange with the best found solutions. The dependence of the quality of the solution on the number of iterations is shown in Figure 4.

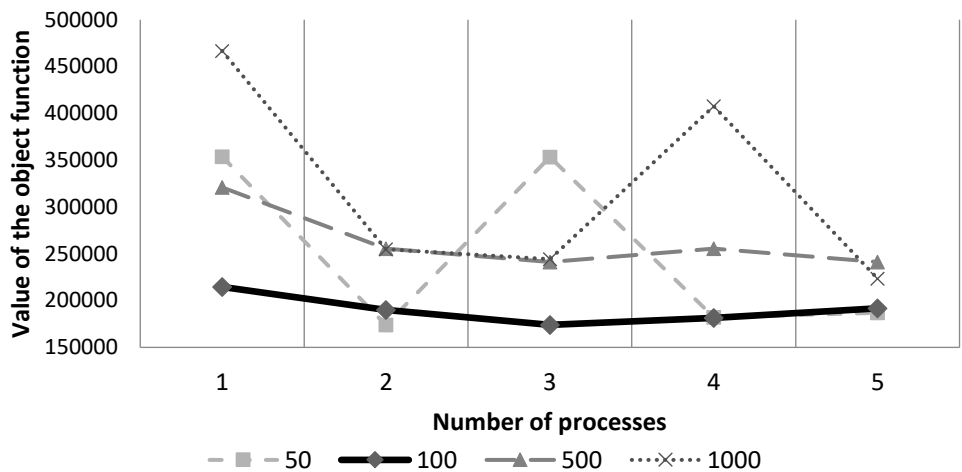

Figure 4. Dependence of objective function values on the number of successive iterations for simulated annealing

You can see from Figure 4 that the best solution quality is achieved with 100 consecutive iterations per one information exchange. When we use other values of this parameter and different orders, the average best solution can be achieved with 1000 consecutive iterations per information exchange. The total number of iterations of the parallel algorithm is related to the number of sequential iterations by the formula: $N = c \times n$, where $N$ is total number of iterations, $c$ is number of information exchanges, $n$ is number of successive iterations.

When we run a fixed number of iterations of the parallel algorithm, decrease in the number of consecutive iterations results in exchanges increase. It affects the execution time of the algorithm.

Depending on the graph order, a different number of iterations of the parallel algorithm may be required. During the selection of parameter values, it was found that:

- if graph contains less than 256 vertices, it is sufficient to run 50,000 iterations of parallel algorithm;
- if graph contains from 256 to 1024 vertices, it is sufficient to run 100,000 iterations of parallel algorithm.

When performing simulated annealing, multiple solvers run simultaneously in each process, trying to improve the current solution. Experiments show that for graphs with no more than 100 vertices, the number of "solvers" coincides with the order of the graph. For graphs containing up to 1024 vertices, there should be 125 solvers. In Figure 5 one can see the dependence of the solutions quality of on the number of solvers for the tai343e01 instance.

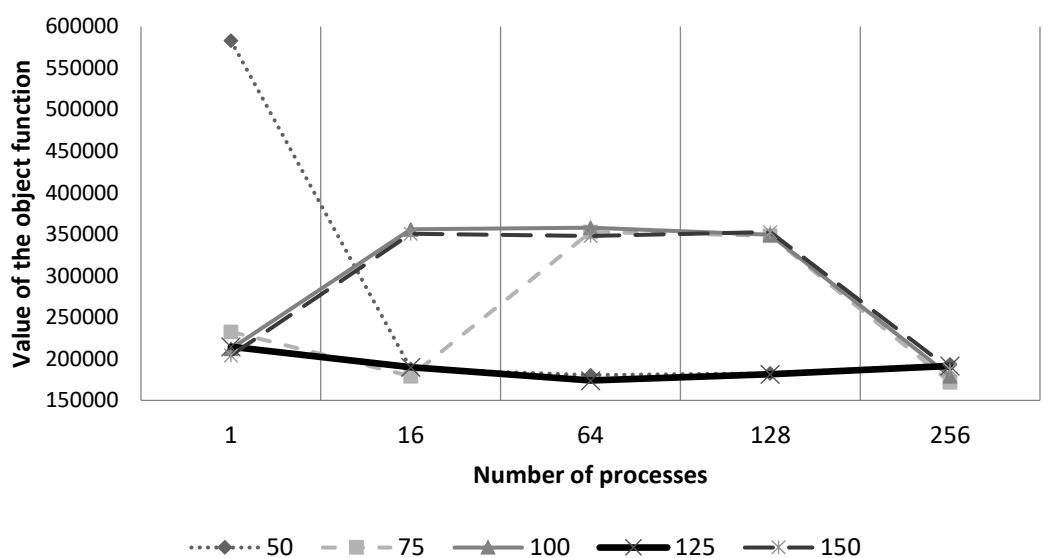

Figure 5. Dependence of the solutions quality of on the number of solvers

During execution of the parallel genetic algorithm, the processes exchange the best individuals after each iteration. The only difference from the sequential algorithm is in information exchanges. For the parallel genetic algorithm there were used the parameters recommended by the author of the UGR-Metaheuristics library, with the exception of the population size:

- probability of crossover is 1, the crossover operation is the basic one, it gives on average the best quality of solutions, compared with crossover with sorting;
- probability of mutation is 0.001;
- the number of migration solutions should be small, as experiments have shown that more than one migration solution degrades the quality of the final solution;
- the number of the population members is equal to the order of the initial graphs;

- for the graphs of different orders, a fixed number of iterations of the algorithm is specified. A fixed number of iterations for the high orders graphs makes it possible to achieve an acceptable solution in a reasonable time.

The composite parallel algorithm consists of a sequential simulated annealing and a parallel genetic algorithms. The solutions of the simulated annealing algorithm are used as the initial population for the parallel genetic algorithm. We set the same parameters for the sequential and parallel simulated annealing algorithms. The parameters of the parallel genetic algorithm were not changed.

At the second stage, the algorithms were compared in terms of accuracy and execution time. Accuracy was measured as the value of the objective function. Note that due to the specific of the parallel algorithms an increase in the number of processes results in expansion of the space of candidate solutions. So, change in the number of processes affects the accuracy of the mapping, almost without affect on the runtime of the algorithms.

Experiments have shown that the parallel genetic algorithm for mapping of large graphs gives on average a more accurate solution than a parallel simulated annealing algorithm. Due to the migration of solutions between different populations, it is possible to transfer the best genes. The parallel genetic algorithm has a longer mapping search time compared to the simulated annealing algorithm for large graphs. This is because for any new descendant, the value of the objective function must be recalculated, in contrast to simulated annealing, where the value of the objective function is calculated relative to the changes made to the mapping.

Dependence of the solution quality on the number of processes for the tai343e01 instance is presented in Figure 6.

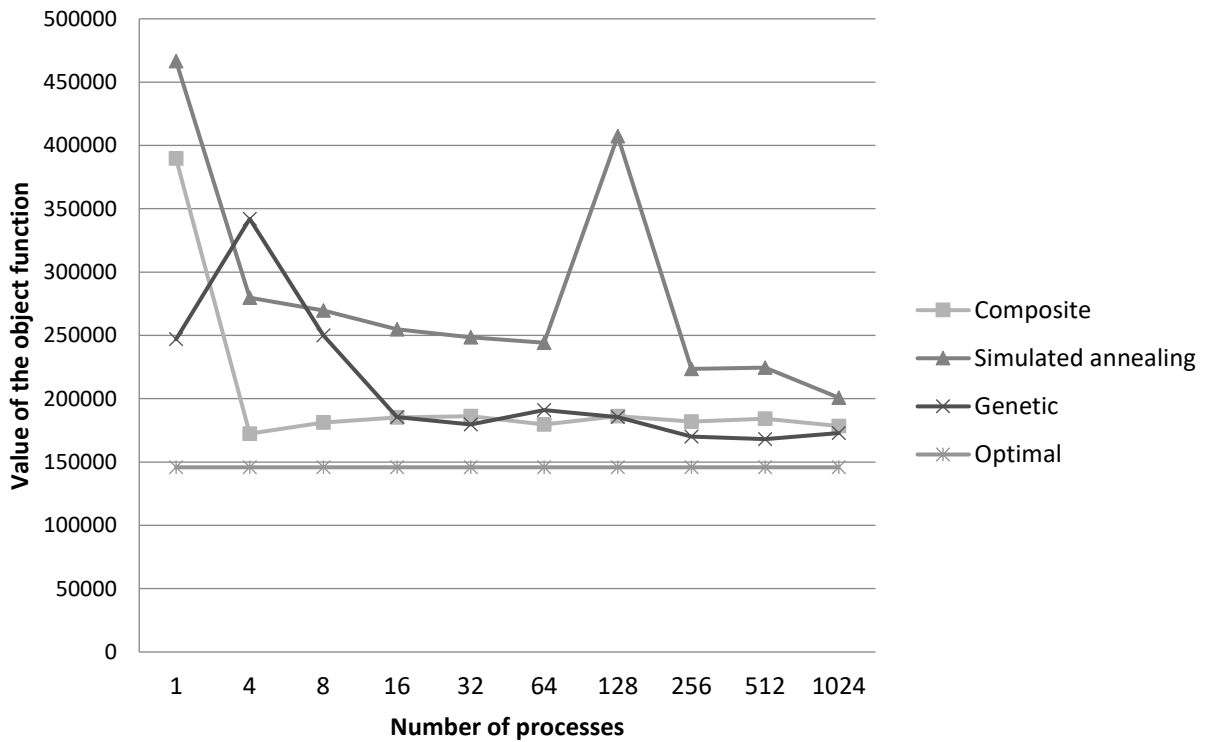

Figure 6. Dependence of the solution quality on the number of processes for the tai343e01 instance

Dependence of the solution quality on number of processes for the tai729e01 instance is shown in Figure 7.

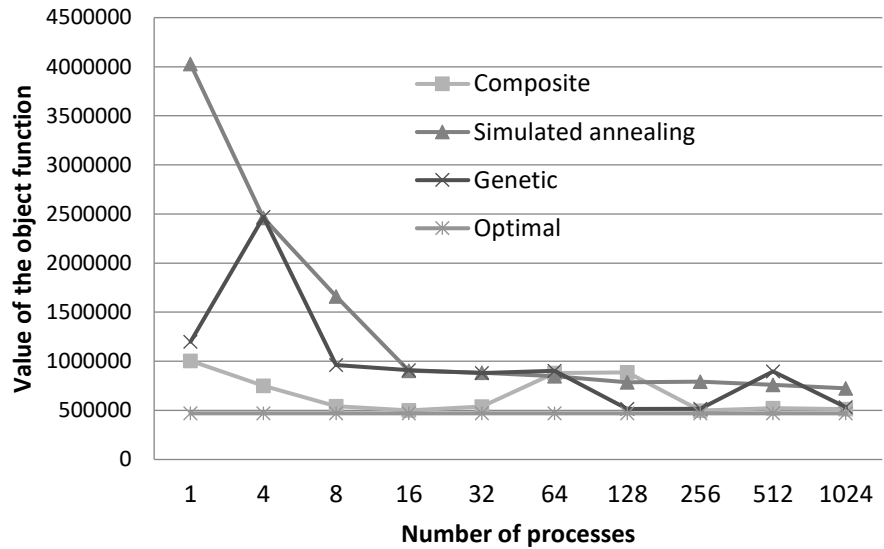

Figure 7. Dependence of the solution quality on number of processes for the tai729e01

Average execution time of algorithms for different data sets is given in Figure 8.

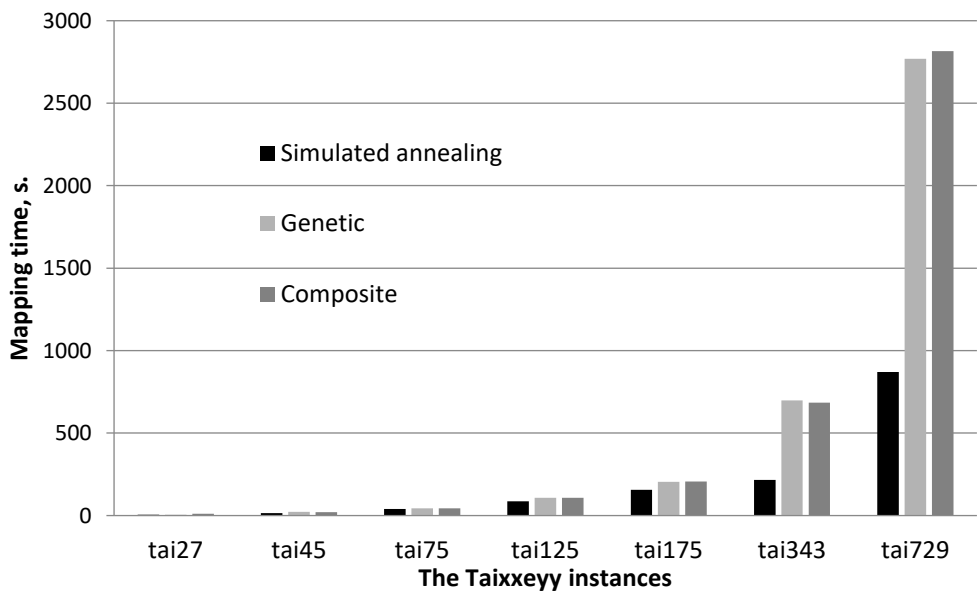

Figure 8. Average execution time of algorithms for different instances

The estimates of the accuracy and execution time according to the experiments of the algorithms are summarized in Table 1. The optimal solution was taken from http://mistic.heig-vd.ch/taillard/problemes.dir/qap.dir/summary_bvk.txt. The column indicates the best experimental value of the objective function. The column indicates the optimal value of the objective function for each instance. The column indicates the time in minutes, which the algorithm worked to get the best solution. The column

12indicates the time to get the optimal mapping from the article [33]. The column contains an estimate of the accuracy of the solution, it shows how the difference of the objective function value from the optimal one in percent: $A_1 = 100 \cdot \frac{F - F_0}{F_0}$.

Table 1. Estimation of the solutions accuracy

| Algorithms<br><br>Taixxeyy instances | Parallel Simulated Annealing | | | Parallel Genetic | | | Parallel Composite | | | Optimal Solution | |
|---|---|---|---|---|---|---|---|---|---|---|---|
| | $F$ | $T$ | $A_1$ | $F$ | $T$ | $A_1$ | $F$ | $T$ | $A_1$ | $F_0$ | $T_0$ |
| Tai27 | 2558 | 0.05 | 1 | 3176 | 0.1 | 24 | 2600 | 0.27 | 2 | 2558 | 0.02 |
| Tai45 | 6724 | 0.3 | 5 | 8564 | 0.45 | 34 | 7332 | 0.5 | 14 | 6412 | 0.03 |
| Tai75 | 19380 | 0.6 | 34 | 18268 | 0.7 | 26 | 18810 | 0.75 | 29 | 14488 | 8 |
| Tai125 | 50780 | 1.6 | 43 | 47816 | 2 | 35 | 50792 | 1.75 | 43 | 35426 | 166 |
| Tai175 | 72688 | 2.8 | 26 | 74602 | 5 | 29 | 74880 | 3.1 | 29 | 57540 | 181 |
| Tai343 | 200856 | 3.5 | 37 | 168120 | 12.8 | 15 | 172466 | 10.1 | 18 | 145862 | 1026 |
| Tai729 | 724820 | 18.2 | 54 | 514846 | 50 | 9 | 498454 | 53.2 | 6 | 469650 | 1187 |

Let us consider applicability of the results to the supercomputer resource managers. The following can be noted.

1. Only simulated annealing shows acceptable times for large graphs. Only this mapping algorithm requires time comparable to system timeouts (no more than 15 minutes). Genetic and composite algorithms find better solutions, but take an unacceptably long time.
2. The two-stage PGA method [2] includes execution of mapping algorithms on the supercomputer nodes allocated for the next in time job. When this method is used, the comparison of the parallel algorithms accuracy must be performed on the number of processes equal to the number of cores (nodes) allocated for the job. Experimental results show that on the number of processes equal to the number of graph vertices, the accuracy of simulated annealing is comparable to the accuracy of the genetic and composite algorithms. The exception is the case of a graph of 729 vertices, for which both the genetic and composite algorithms demonstrate high accuracy with an error of 9% and 6%, respectively.

## 6. CONCLUSION

For quick mapping of a program graph onto a graph of a supercomputer we executed parallel algorithms on the nodes of the supercomputer allocated to the job. We experimentally compared the parallel simulated annealing algorithm, the genetic algorithm, and the composite algorithm.

The parallel combined algorithm always provides an average more accurate solution than the genetic and simulated annealing algorithms. The simulated annealing algorithm has the minimum runtime. The runtime of the composite algorithm is close to the runtime of the parallel genetic algorithm. With a small number of processes the composite algorithm provides acceptable solutions, this reduces the amount of resources for mapping. For high order graphs, the most accurate solutions are found by the parallel genetic algorithm and the parallel composite algorithm.

The parallel simulated annealing algorithm requires significantly less time, and it can be recommended for mapping the regular jobs in supercomputer resource managers.

Acknowledgments. The work was carried out at the JSCC RAS as part of the government assignment FNEF-2022-0014. Supercomputer MVS-10P OP was used.